\documentclass[9pt,sigconf, superscriptaddress]{acmart}
 \renewcommand\footnotetextcopyrightpermission[1]{} 
\usepackage{booktabs} 
\usepackage[utf8]{inputenc}
\usepackage{color}
\usepackage{xcolor}
\usepackage[ruled, vlined, linesnumbered]{algorithm2e}
\usepackage{algpseudocode}
\usepackage{url}
\usepackage{epsfig,endnotes,multirow,multicol}
\usepackage{ifthen}
\usepackage{datetime}
\usepackage{xspace}
\usepackage{todonotes}
\usepackage{graphicx}
\usepackage{caption}
\usepackage{subcaption}
\usepackage[miktex]{gnuplottex}

\algnewcommand\True{\textbf{true}\space}
\algnewcommand\False{\textbf{false}\space}

\begin{document}

\title[Faster Control Plane Experimentation with Horse]{Faster Control Plane Experimentation with Horse}

\author{Eder Le\~{a}o Fernandes \qquad Gianni Antichi \qquad Timm Böttger \qquad Ignacio Castro \qquad Steve Uhlig\\ Queen Mary, University of London \\ \{e.leao, g.antichi, timm.boettger, i.castro, steve.uhlig\}@qmul.ac.uk}

\renewcommand{\shortauthors}{Eder Le\~{a}o Fernandes et al.}





\begin{abstract}
Simulation and emulation are popular approaches for experimentation in Computer Networks. However, due to their respective inherent drawbacks, 
existing solutions cannot perform both fast and realistic control plane experiments. To close this gap, we introduce Horse. Horse is a hybrid solution with an emulated control plane, for realism, and simulated data plane, for speed. Our decoupling of the control and data plane allows us to speed up the experiments without sacrificing control plane realism.
\end{abstract}



\maketitle

\section{Introduction}

Computer Network emulators (e.g., Mininet~\cite{mininet2}) and simulators (e.g., NS3~\cite{ns3}) are the most accessible choices for network experimentation at scale. To select the most appropriate tool, researchers have to consider the inherent trade-offs of emulation and simulation techniques. 

Emulators are more \textit{realistic}, but \textit{scalability is limited}.
Emulators rely on virtual nodes with actual network stacks. 
Accordingly, emulation is constrained by hardware resources undermining experimentation of large networks and traffic volumes within a single machine. While emulators can be distributed to scale further, this adds complexity and costs. On the other hand, simulators tend to be \textit{less realistic but more scalable}. Based on models that represent a network, simulators come in many degrees of abstraction and varying degree of complexity. Unfortunately, the speed of experiments on simulators tends to decrease with growing levels of detail. For example, the reproduction of every single packet interaction by most NS3 models inflicts a huge penalty on execution speed. 

Researchers have already proposed solutions to reduce the drawbacks 
of simulation and emulation in many ways. Fs-sdn~\cite{fs-sdn} improves simulation speed, but its scope is limited to SDN environments. Mininet-VT~\cite{VT-Mininet} and Selena~\cite{selena} improve emulation accuracy with virtual time scheduling approaches, however at the cost of execution speed and changes to the Operating System (OS) kernel. Finally, S3Fnet~\cite{S3Fnet} offers the possibility to mix parallel simulation and emulation. However, it also requires kernel changes and speed suffers under high data plane loads.

Unfortunately none of these tools provides fast experiments with realistic control plane. For this reason we designed and implemented Horse. Contrarily to typical hybrid simulations in which nodes are either fully emulated or simulated, Horse \textit{decouples the network planes} by applying emulation to the control plane and simulation to the data plane. This enables the speed up or slow down of the experiment according to the occurrence of control plane updates. Moreover, the design of Horse does not require changes to the kernel of the OS.

\section{Design and Implementation}
                               
\textbf{Design.} The main premises of Horse are (1) the emulation of the control plane, with actual routing protocols or SDN controllers; and (2) a simplistic simulated data plane that runs a fluid rate traffic model. When control plane events occur, such as a Border Gateway Protocol (BGP) update message or an OpenFlow Flow-Mod message, Horse detects the event and runs (or stays) in a mode named \textit{Fixed Time Increment (FTI)}. The FTI mode reproduces the real-time operation of the control plane by increasing the experiment time in small fixed intervals. After a user-defined timeout without control plane events, Horse advances time as a traditional \textit{Discrete Event Simulator (DES)}, in which the clock is set to the time of the executing event. Note that both the emulated control and the simulated data plane execute at the same time. The key point is the transition of time to adapt to real time when there are emulated control events and the faster advance when only data plane traffic flows in the network. 

\begin{figure}[h!] 
\includegraphics[width=\columnwidth]{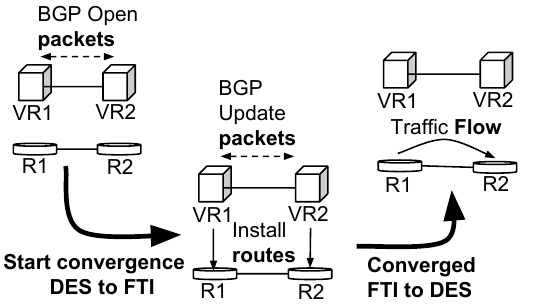} 
\caption{Transition between execution modes in a scenario with two BGP Routers.}   
\label{fig:Routerexample} 
\end{figure}

Figure~\ref{fig:Routerexample} depicts the operation of Horse in a scenario with two BGP routers, R1 and R2. VR1 and VR2 represent the emulated instances of the routers' control planes. The experiment changes from DES to FTI when the routers have established a BGP session. As long as both parties exchange updates, the experiment remains in the FTI mode. When the routers add routes to their Routing Information Base (RIB), Horse installs those routes in the respective data planes. After convergence, no more updates are sent and the experiment resumes to the faster DES mode. Horse also handles SDN networks. In this case the control plane packets are actually sent to the data plane allowing for programmability.

\begin{figure}[h]     
\includegraphics[width=\columnwidth]{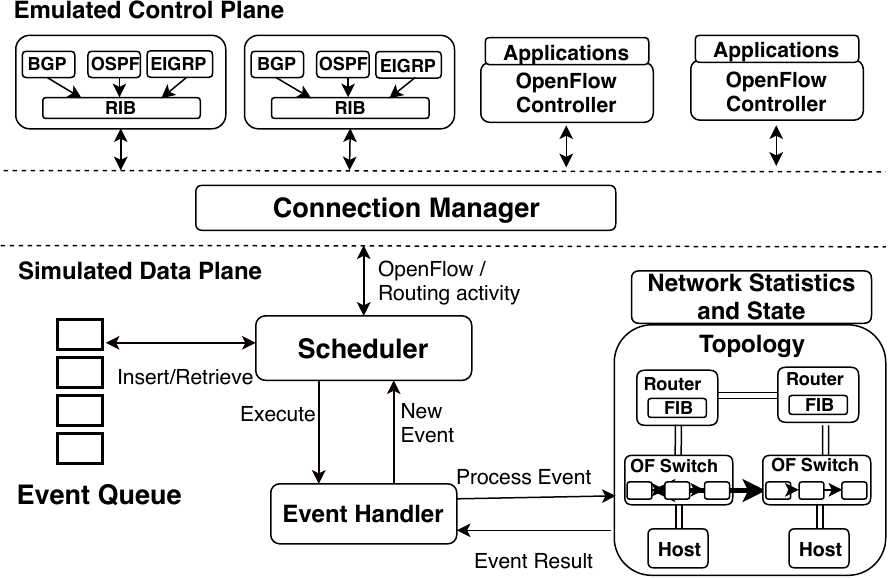}     
\caption{General Architecture of Horse}     
\label{fig:sim_arch} 
\end{figure}

 
\textbf{Implementation.} We developed Horse using the C language with a Python API to make it easier to build and run experiments. The implementation follows the architecture in Figure~\ref{fig:sim_arch}. In the control plane, routers execute real routing daemons and the controllers execute actual SDN applications. The Connection Manager (CM) is the bridge between the emulation and simulation. The CM has visibility to control plane packets and is responsible for sending events that trigger a change to the FTI mode. The data plane is a typical DES engine, with an Event Queue, Scheduler and a simulated model of the nodes of the network topology. The current implementation of Horse supports routers with Quagga as routing daemon and OpenFlow enabled switches. In the future, we plan to also support P4 switches. The code from Horse is Open Source and is available online~\footnote{https://github.com/ederlf/horse}.

\section{Demonstration}

\begin{figure}[h]     
\includegraphics[width=\columnwidth]{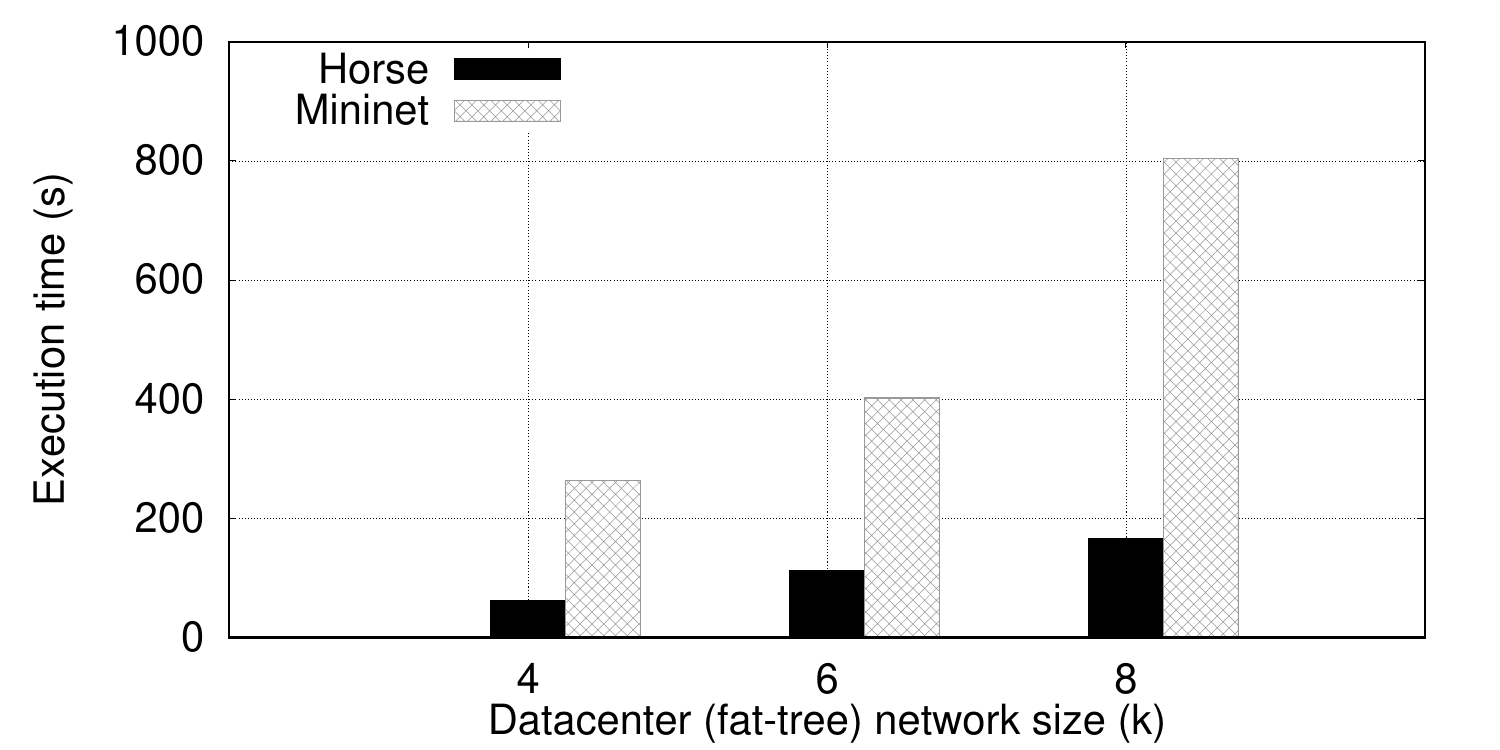}     
\caption{Execution time of the demonstration on Horse and Mininet}     
\label{fig:time} 
\end{figure}

The demonstration presents how Horse enables quicker experimentation with BGP and SDN control planes. We use a Fat-Tree topology~\cite{fattree} to demonstrate the time required to perform experiments on different network sizes. Even though the demonstration focus on a Data Center (DC) scenario, Horse is not restricted to DCs and can also be used for other types of networks, e.g., Wide Area Networks (WAN). The demonstration consists of three experiments showcasing different Traffic Engineering (TE) approaches to achieve better link utilization: (i) BGP plus Equal Cost Multipath (ECMP) path selection by hashing of IP source and destination; (ii) Hedera~\cite{alfares10}; (iii) SDN 5-tuple (IP source and destination, IP protocol, transport source and destination ports) ECMP. The Fat-Trees for the scenarios have $4$, $6$ and $8$ pods with links of $1Gbps$. A single traffic pattern is used for all experiments: each server of the DC sends a single UDP flow to another server inside the DC, at the constant rate of $1Gbps$. If no congestion occurred, the total traffic rate expected in the network would be equal to the number of hosts (For example, for $4$ with $16$ hosts, the total traffic is $16Gbps$). The execution of the experiments for each topology starts simultaneously. For each experiment, we show the amount of time required to create the topology and the consolidated time to execute the three TE approaches. At the end of each execution, we show a graph of the aggregated rate of all flows arriving at the hosts for each TE case. These three TE approaches are chosen because of different levels of control plane interactions. The cases of ECMP for SDN and BGP, control plane events are concentrated at the beginning of the simulation while our implementation of Hedera queries for network statistics every $5$ seconds. The demonstration of different scenarios is meant to help researchers to understand how the tool can be used and whether it is a good match for their use cases. For comparison, it would be interesting to demonstrate Horse along with Mininet. However, as shown by Figure~\ref{fig:time}, Mininet takes  5 times longer than Horse to finish the largest topology~\footnote{Experiments performed in a virtual machine with 4GB of RAM with four cores of an Intel(R) Xeon(R) Silver 4114 @ 2.20GHz CPU assigned.}. The difference of time would break the flow of the demonstration.

\section*{Acknowledgments}

This research is supported by the UK's Engineering and Physical Sciences 
Research Council (EPSRC) under the EARL: sdn EnAbled MeasuRement for alL project (Project Reference EP/P025374/1). 

\label{ConcPage}
\bibliographystyle{ACM-Reference-Format}
\bibliography{paper}







\end{document}